\documentclass[showpacs,amsmath,amssymb,pra,superscriptaddress,preprint]{revtex4-1}
\usepackage{graphicx}
\usepackage{amsmath}
\usepackage{hyperref}
\usepackage{dcolumn}
\usepackage{bm}

\begin{document}

\title{Field theoretical approach to quantum transmission in time-dependent potentials}

\author{F.K. Diakonos}
\email[]{fdiakono@phys.uoa.gr}
\affiliation{Department of Physics, University of Athens, GR-15771 Athens, Greece}
\author{P.A. Kalozoumis}
\email[]{pkalozoum@phys.uoa.gr} \affiliation{Department of
Physics, University of Athens, GR-15771 Athens, Greece}
\author{A.I. Karanikas}
\email[]{akaran@phys.uoa.gr}
\affiliation{Department of Physics, University of Athens, GR-15771 Athens, Greece}
\author{N. Manifavas}
\email[]{nmanif@phys.uoa.gr}
\affiliation{Department of Physics, University of Athens, GR-15771 Athens, Greece}
\author{P. Schmelcher}
\email[]{pschmelc@physnet.uni-hamburg.de}
\affiliation{Zentrum fuer Optische Quantentechnologien, Universit\"{a}t Hamburg,
Luruper Chaussee 149, 22761 Hamburg, Germany}
\date{\today}

\begin{abstract}
We develop a field theoretical approach based on the temporary basis description as a tool to investigate the transmission properties of a time-driven quantum device.
It employs a perturbative scheme for the calculation of the transmission of a monochromatic beam of particles through the time-dependent set-up. The main advantage of the proposed treatment is that it permits the use of the particle picture for
the calculation of the scattering matrix and the transmission coefficient. Therefore the elementary physical processes contributing to the transmission can be identified and interpreted in a transparent way. We apply the method to the simple but prototype problem of transmission through an one-dimensional oscillating delta potential and we demonstrate how it enables a deep understanding of the underlying physical processes.
\end{abstract}
\maketitle

\section{Introduction}

The transmission of quantum particles through a time-dependent
potential has been the subject of extensive studies in the last
three decades (see \cite{delCampo2009} and references therein).
The main goal in these studies is to classify and understand the
mechanisms of quantum tunnelling in such a potential. Typical
examples of processes involving transmission through time
dependent devices are the tunnelling of a test particle from a
metastable state \cite{Caldeira1983} using instanton techniques, the scattering off a
dissipative environment \cite{Bruinsma1986} and the tunneling through a time-modulated
barrier \cite{Buttiker1982}. Particularly in \cite{Buttiker1982} it was
shown that at low modulation frequencies the traversing particle
sees a static barrier and at high frequencies the particle tunnels
through the time-averaged potential. It was also demonstrated that
inelastic processes can occur, where the tunneling particle gains
or loses energy quanta from the modulation field. In the same work
a fundamental question was raised concerning the definition
of the traversal time through such a fluctuating device, an issue
which is still under debate \cite{Hauge1989}. Recently novel 
perspectives for the study of driven quantum systems emerged. 
In particular it has been
realized that there is the possibility to exploit phenomena occurring in
these systems for the development of novel technology such as the design of
quantum pumps, i.e. devices capable to create quantum directed
transport through periodic external driving \cite{Switkes1999}, or
the control of quantum devices for information processing
\cite{Miller1987}.

One of the simplest examples for studying the transmission in a
time-dependent environment is the tunnelling through a
delta-barrier with a harmonically oscillating coupling which has been considered by
several authors
\cite{Pimpale1991, Bagwell1992, Martinez2001}.
Although the device is very simple, the presence of the
time-dependent coupling leads to interesting phenomena, observable
in the transmission coefficient, such as Fano resonances, threshold
enhancements related to sideband modes and their interplay
\cite{Bagwell1992}. More complicated setups, involving several
oscillating delta barriers, have been investigated in the context
of quantum pumps \cite{Moskalets2003}. In this case dephasing
effects may also occur, influencing the appearance of resonances
\cite{Diakonos2011}. In \cite{Martinez2002} an infinite periodic
chain of delta-barriers with harmonically oscillating strength has
been studied demonstrating the modification of the conductance
zones through the time-dependence.

If the driving of the potential barrier is periodic the standard
approach used to solve the quantum dynamics is based on Floquet
theory \cite{Tannor2007} which is very well suited for the straightforward
calculation of the transmission properties, discriminating between the
contributions from elastic and inelastic channels. In the Floquet treatment
the considered problem is first rendered time-independent and then solved
using standard numerical techniques.
This represents certainly an advantage from the computational point of view. However, it is
conceptually a handicap since it does not allow to isolate contributions
or identify mechanisms
leading to a specific dynamical behaviour that allow for a better understanding
of the underlying physics.

Path integral methods, on the other hand side, have been commonly used for the study
of the quantum dynamics in time-dependent potentials
\cite{Kleinert2009} but they show usually a slow convergence. Thus,
despite of a few exceptional cases where an analytical solution
can be obtained \cite{Storchak1992}, an efficient general purpose
path integral treatment of time-dependent potentials is still
lacking. Progress in this direction has recently been achieved by
introducing a rapidly converging scheme in the framework of a high-order
short-time expansion of transition amplitudes in time-dependent
potentials  \cite{Balaz2011}. This approach holds for a general time-dependence 
of the potential. The disadvantage of this approach is that it can only be applied
to smooth potentials. This is a relevant point in view of the fact that 
previous works have mostly been employing either the $\delta$ or a 
rectangular barrier for the study of the time-dependent systems.

In the present work we develop a field theoretical
approach to the investigation of the quantum
dynamics in time-dependent potentials. The method is suitable for
studying any kind of potentials (also non-smooth) as well as
external driving (periodic or not). In addition it allows for a 
classification and decomposition of the quantum dynamics 
in terms of fundamental processes
which represent the skeleton of the quantum evolution in these
systems. Particularly we reveal the role of virtual ``multi-photon" processes which
determine to a large extent details of the resonant structures in the dependence of the 
transmission coefficient on the incoming energy. Within our treatment these sub-processes 
can be clearly distinguished from the real ``multi-photon" exchange which lead to inelastic 
transmission \cite{Buttiker1982}. 
The power of our method is demonstrated using the example
of the oscillating delta barrier which has been extensively studied in
the literature. We show that it is possible to calculate and
understand transmission properties which have not been accessible
so far. An example is the transmission zero associated with the Fano resonance characterizing 
the quantum dynamics in this system. The virtual ``multi-photon" processes, mentioned above, 
are relevant for the determination of its location.

The paper is organized as follows. In section 2 we present
the main idea of the perturbative scheme used to describe the
transmission properties of a time-dependent quantum device. In
section 3 we present our method focusing on
the case of periodic driving and using as a simple example the
delta barrier with a harmonically oscillating strength. In section
4 we give the results for the transmission properties of the
oscillating delta-barrier setup. In particular we demonstrate how
one can systematically calculate and interpret details of the
behaviour of the transmission coefficient as a function of the
incoming energy within our approach. Section 5 contains our
concluding remarks. Finally, extensive formulas and their
derivation are provided in the appendices.

\section{Field theoretical perturbative approach for transmission in time-dependent potentials}

We consider the transmission of a quantum particle through a
localized time-dependent potential. Our specific analysis is performed
assuming a one-dimensional setup. The proposed approach
can however be easily generalized to higher dimensional cases.
Initially, i.e. for $t \to -\infty$, the wave function of the quantum
particle for $x \to -\infty$ is a plane wave with energy $E_i$ and
momentum directed from the left to the right (incoming quantum
particle). We also assume that the wave function for $t \to
\infty$ and $x \to \infty$ is a plane wave with energy $E_f$ and
the same direction of the momentum (outgoing quantum particle). The
amplitude for the scattering of the considered quantum particle by
a specific local time-dependent potential is in general given in
terms of the corresponding causal Green's function $\Delta_{n,m}$
as:
\begin{equation}
S_{fi}=-i \hbar e^{i(E_f t_f - E_i t_i)/\hbar} \sum_{n,m}  \langle
\vec{p}_f \vert n(t_f) \rangle \Delta_{n,m}(t_f,t_i) \langle
m(t_i) \vert \vec{p}_i \rangle \label{eq:eq1}
\end{equation}
where $\vec{p}_i$ and $\vec{p}_f$ are the input and output momenta
of the particle while $E_i$ and $E_f$ are the corresponding
energies. In eq.~(\ref{eq:eq1}) we have introduced the temporary
basis $\{ \vert n(t) \rangle \}$ with $n$ being a collective index for
all the quantum numbers needed to fully identify each temporary
eigenstate of the time-dependent Hamiltonian $\hat{H}(t)$ at time $t$. In this
basis the instantaneous Schr\"{o}dinger equation reads:
\begin{equation}
\hat{H}(t)\vert n(t) \rangle = E_n(t) \vert n(t) \rangle
\label{eq:eq2}
\end{equation}
The relevant quantity for determining the Green's function $\Delta_{n,m}$ are the geometrical phases
$\langle n(t) \vert i \hbar \partial_t \vert m(t) \rangle$ given as:
\begin{equation}
\langle n(t) \vert i \hbar \partial_t \vert m(t) \rangle=
\frac{\langle n(t) \vert i \hbar \partial_t \hat{H}(t) \vert m(t)
\rangle}{E_m(t)-E_n(t)}~~~~~~,~~~~~ (m \neq n) \label{eq:eq3}
\end{equation}
The diagonal elements $\gamma_n(t) \equiv \langle n(t)  \vert i
\hbar \partial_t \vert n(t) \rangle$ are the Berry phases. In the
temporary basis the propagating kernel $\Delta_{n,m}$ fulfils the
Green's equation:
\begin{equation}
\sum_{m} [(i \hbar \partial_t - \tilde{E}_n(t))\delta_{n,m} -
\Phi_{n,m}(t)]\Delta_{m,n'}(t,t')=-\delta_{n,n'}\delta(t-t')
\label{eq:eq4}
\end{equation}
where the infinite dimensional matrix $\Phi$ contains only the
non-diagonal geometrical phases:
\begin{equation}
\Phi_{n,m}(t)=\langle n(t) \vert i \hbar \partial_t  \vert m(t)
\rangle - \delta_{n,m} \gamma_m(t) \label{eq:eq5}
\end{equation}
while the Berry phases are included in the effective energies $\tilde{E}_n(t)$:
\begin{equation}
\tilde{E}_{n}(t)=E_n(t)-\sum_{m} \delta_{n,m} \gamma_m(t)
\label{eq:eq6}
\end{equation}
In operator notation the propagator $\Delta_{n,m}$ can be written as:
\begin{equation}
\Delta_{n,m}(t_f,t_i)=\langle n(t_f) \vert \frac{1}{\hat{H}_0
-\tilde{E} - \hat{\Phi}} \vert m(t_i)
\rangle~~~~~~;~~~~~~\hat{H}_0=-i \hbar \partial_t \label{eq:eq7}
\end{equation}
The non-diagonal matrix $\Phi$ in (\ref{eq:eq7}) renders the
considered problem analytically intractable in its general form.
To proceed with the calculation of the amplitude (\ref{eq:eq1}) it
is necessary to develop a scheme allowing for the expansion of the
propagator in terms of simpler calculable sub-processes. The main
assumption in our approach is that there exists an ordering with respect to the
magnitude of the transition amplitudes of the different dynamical processes
taking place in the scattering off a time-dependent potential. The latter is
implied by the energy difference between the incoming and the
outgoing state as stated in eq.(\ref{eq:eq3}).  The amplitudes of
the elastic processes dominate while inelastic processes with
small energy transfer are more probable than those with large
energy transfer. This property, if valid and consistently applicable, suggests that the
non-diagonal matrix $\Phi$ can be treated as a perturbation and
the following expansion of $\Delta_{n,m}$ is possible:
\begin{equation}
\Delta \equiv \frac{1}{\hat{H}_0 - \tilde{E} - \hat{\Phi}} = 
\frac{1}{\hat{H}_0 - \tilde{E}} + \frac{1}{\hat{H}_0 - \tilde{E}}
\hat{\Phi} \frac{1}{\hat{H}_0 - \tilde{E}} + ... \label{eq:eq8}
\end{equation}
In eq.~(\ref{eq:eq8}) the expansion breaks down in the case of
zero  eigenvalues of the denominator $\hat{H}_0 - \tilde{E}$
requiring a special treatment. We will come back to this point
later on. Using the expansion (\ref{eq:eq8}) we can write the
amplitude (\ref{eq:eq1}) as follows:
\begin{equation}
S_{fi}=S_{fi,static} - i \hbar e^{i(E_f t_f - E_i t_i)/\hbar}
\sum_{n,m} \langle \vec{p}_f \vert n(t_f) \rangle
\sum_{r=1}^{\infty} \Delta^{(r)}_{n,m}(t_f,t_i)\langle m(t_i)
\vert \vec{p}_i \rangle \label{eq:eq9}
\end{equation}
where the $\Delta^{(i)}_{n,m}$ term in the above sum involves $i$
insertions of the transition matrix $\Phi$ while $S_{fi,static}$
is the zeroth order term which does not contain $\hat{\Phi}$. The
first term in the sum on the r.h.s. of eq.~(\ref{eq:eq9}) is:
\begin{equation}
\Delta^{(1)}_{n,m}(t_f,t_i)=\int_{-\infty}^{+\infty} dt_1
\sum_{l_1,l_2} \langle n(t_f) \vert
\frac{1}{\hat{H}_0 - \tilde{E} } \vert l_1(t_1) \rangle
\Phi_{l_1,l_2} \langle l_2(t_1) \vert  \frac{1}{\hat{H}_0 -
\tilde{E} } \vert m(t_i) \rangle \label{eq:eq10}
\end{equation}
and the higher order terms have a similar structure as implied by
the expansion (\ref{eq:eq8}). The form (\ref{eq:eq9}) allows the
use of a particle picture for the interpretation of the dynamics
in scattering off a time-dependent potential having at the same
time a simple diagrammatic interpretation: the amplitude $S_{fi}$
is decomposed in a sum of sub-processes. Each sub-process is a
sequence of two elementary processes: the particle propagation
being in a particular eigenstate and the transition between two
states of the temporary basis. The explicit form of $S_{fi}$
depends on the applied potential and especially on the temporary
basis. The power of the diagrammatic representation of $S_{fi}$ is
that it allows the calculation of desired properties (like the
energy of a transmission zero or a local transmission maximum) by
isolating the contributing dynamical processes. In general the
spectrum of $\hat{H}(t)$ contains both a discrete as well as a
continuous part. Therefore the allowed elementary processes can be
classified as follows:
\begin{itemize}
\item continuum-continuum (c/c) transitions
\item continuum-bound (c/b) and bound-continuum (b/c) transitions
\item bound-bound (b/b) transitions
\item propagation in a continuum state
\item propagation in the bound state
\end{itemize}
A typical sub-process contributing to $S_{fi}$ is shown in Fig.~1.
The curly line indicates propagation in a continuum state while
the dashed line means propagation in a bound state. The full black circles
indicate c/c transitions while the circles
containing a cross indicate a c/b or b/c transition.

\begin{figure}[htbp]
\centerline{\includegraphics[width=13 cm,height=1
cm]{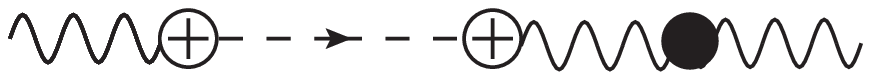}} \caption{A typical sub-process involving 3 transitions
(continuum - bound state,  bound state - continuum and continuum -
continuum) contributing to the transition amplitude (1) at the
third order of the proposed perturbation scheme.} \label{fig:1}
\end{figure}

\section{The case of time-periodic potentials: a delta-barrier with oscillating strength}

The calculation of the transmission properties in a time-dependent
potential  significantly simplifies if the driving is periodic. In
this section we will demonstrate how the perturbative scheme
introduced in the previous section works in practice by performing
an analysis of the transmission properties of a monochromatic wave
passing through a periodically varying potential. As a concrete
example we consider the transmission through a delta barrier with
oscillating strength in one dimension. As mentioned already in the
introduction, our main purpose is to calculate and explain
features of the transmission behaviour which are not easily
accessible by other approaches like direct integration or Floquet
theory. The Schr\"{o}dinger equation of the considered problem
reads:
\begin{equation}
i \hbar \frac{\partial \Psi(x,t)}{\partial t} =
-\frac{\hbar^2}{2m}  \frac{\partial^2 \Psi(x,t)}{\partial x^2} -
g(t) \delta(x) \Psi(x,t)~~~~~~;~~~~~~~g(t)=g(t+T) \label{eq:eq11}
\end{equation}
with $T=\frac{2 \pi}{\omega}$ being the period of the oscillating
$\delta$-potential and $\omega$ the associated frequency.
For fixed $t$ the problem becomes static and can be easily solved.
Introducing the length scale $\ell_0 = \sqrt{\frac{\hbar}{m \omega}}$ 
and the energy scale $\epsilon_0 = \hbar \omega$
the static version of eq.(\ref{eq:eq11}) can be written in
dimensionless form as follows:
\begin{equation}
\left( -\frac{1}{2}\frac{d^2}{d \xi^2} - g_{\tau} \delta(\xi)
\right)  u(\xi;g_{\tau}) = \epsilon(\tau)
u(\xi;g_{\tau})~~~~~~~~~;~~~~~~~\tau=\omega t \label{eq:eq12}
\end{equation}
with $\xi=\frac{x}{\ell_0}$, $g_{\tau}=\frac{1}{\hbar \omega}
\sqrt{\frac{m \omega}{\hbar}} g( \tau)$,
$\epsilon(\tau)=\frac{E(\tau)}{\epsilon_0}$ and $u(\xi;g_{\tau})$
is the corresponding wave function. For a given time instant the
spectrum of the Hamiltonian consists of continuum states and one
bound state with energy $\epsilon_b=-\frac{1}{2}g_{\tau}^2$. The
later exists only during the time period for which $g_{\tau} > 0$.
The associated complete and orthonormal temporary basis is:
\begin{eqnarray}
u^{\pm}_k(\xi;g_{\tau})&=&\frac{1}{\sqrt{2 \pi}}(e^{\pm i k \xi} -
\frac{g_{\tau}}{g_{\tau} + i k} e^{i k \vert \xi \vert})~~~~~~~~k = \sqrt{2 \epsilon}~~~~~\epsilon > 0 \nonumber \\
u_b(\xi;g_{\tau})&=&\sqrt{g_{\tau}} e^{-g_{\tau} \vert \xi \vert} ~~~~~~~~~g_{\tau} > 0
\label{eq:eq13}
\end{eqnarray}
as given in \cite{Campbell2009}.

The perturbative calculation of the scattering amplitude
(eq.~(\ref{eq:eq1}))  requires the summation of all the
contributing sub-processes in increasing order. According to the
discussion in the previous section the order of a term in the
perturbation series is determined by the number of transitions
(c/c, b/c or c/b)  and can
be diagrammatically presented by the series of graphs shown up to
second order in Fig.~2. Since there is only one bound state in the
temporary spectrum there are no possible transitions between bound
states.
\vspace*{0.5cm}
\begin{figure}[htbp]
\centerline{\includegraphics[width=17 cm,height=0.7
cm]{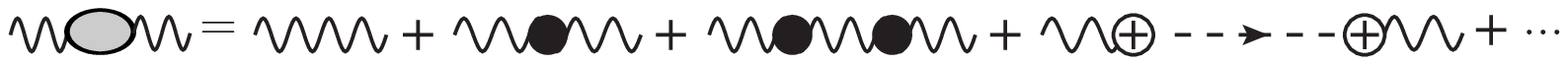}} \caption{The diagrammatic expansion for
amplitude (1) up to the second  order in the introduced
perturbative scheme.  } \label{fig:2}
\end{figure}

To first order only processes with a single transition, necessarily of c/c
type,  participate. These may be elastic or inelastic. In
the inelastic case the final energy is given as:
$\epsilon_f=\sqrt{\epsilon_i^2 + n}$ where $n$ takes the values
$n=\pm 1, \pm 2, ..$ under the restriction that $\epsilon_i^2 + n$
remains positive. This is a consequence of the periodic driving
\cite{Buttiker1982} valid for all orders of the proposed
perturbation expansion (real ``multi-photon" exchange).
Before going on with the explicit calculation of the transmission
amplitude  (\ref{eq:eq1}) it is useful to present the matrix elements
for the building blocks of the contributing sub-processes.\\

\noindent
{\it Continuum-continuum transitions}\\
The transition amplitude from one continuum state to another has the form:
\begin{equation}
\Phi_{k k^{\prime}}(\tau) \equiv \frac{1}{\hbar \omega} \langle
k^{(\pm)}(\tau)  \vert i \hbar \partial_{\tau} \vert
{k^{\prime}}^{(\pm)}(\tau) \rangle = \frac{i}{\pi} \dot{g}_{\tau}
\frac{e^{i(\theta_{k^{\prime}}(\tau)-\theta_k(\tau))}}{(g_{\tau}^2+k^2)^{1/2}
(g_{\tau}^2+{k^{\prime}}^2)^{1/2}}\frac{k k^{\prime}}{k^2 -
(k^{\prime} + i \eta)^2} \label{eq:eq14}
\end{equation}
with $\theta_k(\tau)=\arctan(\frac{g_{\tau}}{k})$. The superscript $(\pm)$ is omitted in $\Phi_{k k^{\prime}}(\tau)$ since the r.h.s. of eq.~(\ref{eq:eq14}) is independent of these signs. As expected the
diagonal term $k=k^{\prime}$  diverges and needs to be
regularized. We adopt the usual {\it box regularization} in order
to define a regularized Berry phase:
\begin{equation}
\langle k^{(\pm)}(\tau) \vert i \hbar \partial_{\tau} \vert
k^{(\pm)}(\tau) \rangle_{reg}\equiv \frac{2 \pi}{V} \langle
k^{(\pm)}(\tau) \vert i \hbar \partial_{\tau} \vert
k^{(\pm)}(\tau) \rangle=\frac{1}{\hbar \omega} \gamma_k(\tau)=-
\frac{1}{\hbar \omega}\dot{g}_{\tau} \frac{k}{g_{\tau}^2 + k^2}
\label{eq:eq15}
\end{equation}
where $\frac{V}{2 \pi} = \int_V \frac{d \xi}{2 \pi} = \delta_R(0)$
with $\delta_R(0)$ being the delta function regularized on a finite volume $V$.
With this choice $\eta$ in eq.~(\ref{eq:eq14}) is finite fulfilling $\eta V \varpropto 1$. After including also the off diagonal terms we
get the general, regularized expression, for $\Phi_{k
k^{\prime}}(\tau)$:
\begin{equation}
\Phi_{k k^{\prime}}(\tau)=\frac{1}{\hbar \omega}  \left( \langle
k^{(\pm)}(\tau) \vert i \hbar \partial_{\tau} \vert
{k^{\prime}}^{(\pm)}(\tau) \rangle - \gamma_k(\tau)
\delta(k-k^{\prime}) \right) \label{eq:eq16}
\end{equation}
with $\delta(k-k^{\prime}) \Phi_{k k^{\prime}}(\tau)=0$.\\

\noindent
{\it Continuum-bound state transitions}\\
Denoting with $\vert b(\tau) \rangle$ the unique bound  state of
the potential (existing only for $\tau \in (2 \pi m, 2 \pi m +
\pi)$ with $m=0, \pm 1, \pm 2, ...$) we can express the
c/b transition as:
\begin{equation}
\Phi_{b k}(\tau)=\frac{1}{\hbar \omega} \langle b(\tau)  \vert i
\hbar \partial_{\tau} \vert k^{(\pm)}(\tau) \rangle = -2 i k
\sqrt{\frac{g_{\tau}}{2 \pi}}
\frac{\dot{g}_{\tau}}{(g_{\tau}^2+k^2)^{3/2}} e^{i \theta_k(\tau)}
\label{eq:eq17}
\end{equation}
Obviously the b/c transition amplitude  $\Phi_{k
b}$ is the complex conjugate of $\Phi_{b k}$.

\noindent
The zero-order approximation of the scattering amplitude (\ref{eq:eq1}) is given as:
\begin{eqnarray}
S_{fi}^{(0)}=-i e^{i (\epsilon_f \tau_f -\epsilon_i \tau_i)} [ \Delta_{b}^{(0)}(\tau_f,\tau_i) \langle k_f \vert b\rangle \langle b \vert k_i \rangle - \nonumber \\
 \int_{0}^{\infty} dk \Delta_{k}^{(0)}(\tau_f,\tau_i) 
\left( \langle k_f \vert k^{(+)}(\tau_f) \rangle \langle k^{(+)}(\tau_i) \vert k_i \rangle~+~(+ \leftrightarrow -) \right) ]
\label{eq:eq18}
\end{eqnarray}
In eq.(\ref{eq:eq18}) $\Delta_{b}^{(0)}$ is the propagator in the bound state while $\Delta_{k}^{(0)}$ is the propagator in a continuum state. The causal form of the latter reads:
\begin{equation}
\Delta_{k}^{(0)}(\tau_f,\tau_i)=i \theta(\tau_f-\tau_i) e^{-i \int_{\tau_i}^{\tau_f} d\tau_1 [\epsilon_k -2 \gamma_k(\tau_1)]}
\label{eq:eq19}
\end{equation}
The bound-state propagator is, in the interval $0< \tau, \tau^{\prime} < 2 \pi$, the solution of the Green's equation:
\begin{equation}
\left[ i \partial_{\tau} - \epsilon_b(\tau) \right] \Delta_{b}^{(0)}(\tau,\tau^{\prime})=-\delta(\tau - \tau^{\prime})
\label{eq:eq20}
\end{equation}
The causal solution of eq.(\ref{eq:eq20}) is:
\begin{equation}
\Delta_{b}^{(0)}(\tau,\tau^{\prime})=i \theta(\tau-\tau^{\prime}) e^{-i \int_{\tau^{\prime}}^{\tau} d\tau_1 \epsilon_b(\tau_1) }
\label{eq:eq21}
\end{equation}
while the solution obeying the periodic boundary condition $\Delta^{(0)}_{b}(\tau,\tau^{\prime})=
\Delta^{(0)}_{b}(\tau + 2 \pi n,\tau^{\prime} )~~~(n=0,\pm 1, \pm 2, ..)$ reads:
\begin{equation}
\Delta_{b}^{(0)}(\tau,\tau^{\prime})= i \sum_{m=-\infty}^{\infty} \theta(\tau - \tau^{\prime} - 2 \pi m) e^{-i \int_{\tau^{\prime} + 2 \pi m}^{\tau} d\tau_1 \epsilon_b(\tau_1) }
\label{eq:eq22}
\end{equation}
In obtaining the $S$-matrix amplitude one has to perform the limits $\tau_i \to -\infty$, $\tau_f \to \infty$. For simplicity we will assume here $g(\tau)=g_0 \sin \tau$. Thus a consistent treatment requires $\tau_f=-\tau_i=\lim_{N \to \infty} \frac{N \pi}{\omega}$ such that $g(\tau_f)=g(\tau_i)=0$. In this case the zero-order contribution to the scattering amplitude is trivial containing no transitions (free transmission). We will use in the following the notation $(nt,mc,lb)$ for the classification of
the contribution of the various sub-processes to the $S$-matrix. $nt$ means that the considered sub-process contains in total $n$ transitions. From these $n$ transitions $m$ are of c/c type while $l$ are of b/c or c/b type. Using this notation we write:
\begin{equation} 
S_{fi}^{(0t,0c,0b)}=\delta(k_f-k_i).
\label{eq:eq23}
\end{equation}
The first order term, containing a single c/c transition, becomes:
\begin{equation}
S^{(1t,1c,0b)}_{fi}= -i e^{i (\epsilon_f \tau_f - \epsilon_i \tau_i)}
\int_{-\infty}^{\infty} d\tau_1 \Delta^{(0)}_{k_f}(\tau_f,\tau_1)
\Phi_{k_f k_i}(\tau_1) \Delta^{(0)}_{k_i}(\tau_1,\tau_i)
\label{eq:eq24}
\end{equation}
Obviously there are no first order processes involving a single 
b/c or c/b transition since the outgoing and incoming states
belong necessarily to the continuum spectrum (positive energy).
Eq.~(\ref{eq:eq24}) can be rewritten as:
\begin{equation}
S_{fi}^{(1t,1c,0b)}=2 \pi i \sum_{n \neq 0} \frac{A_{k_f k_i}(n)} {\vert
k_f \vert} \delta(k_f - \sqrt{k_i^2 + 2 n}) + (k_f \to - k_f)
\label{eq:eq25}
\end{equation}
with:
\begin{displaymath}
\displaystyle e^{2i \theta_{k_f}(\tau)} \Phi_{k_f k_i} e^{-2i \theta_{k_i}(\tau)} =
\sum_{n=-\infty}^{\infty} A_{k_f k_i}(n) e^{-i n \tau}  
\end{displaymath}
and therefore
\begin{equation}
A_{k_f k_i}(n)=\int_0^{2 \pi} \frac{d \tau}{2 \pi}
\left[e^{2i \theta_{k_f}(\tau)} \Phi_{k_f k_i} e^{-2i \theta_{k_i}(\tau)} \right] e^{i n \tau}
\label{eq:eq26}
\end{equation}
Since the renormalized c/c transition obeys:
$$\Phi_{k_f k_i} \delta(k_f-k_i)=0$$
the term $n=0$ is not included in the sum of eq.~(\ref{eq:eq25}).
The calculation of the amplitude $A_{k_f k_i}$, introduced to
describe the $S$-matrix contribution of the c/c
transitions, is straightforward and leads to the expression:
\begin{eqnarray}
A_{k_f k_i}(n)&=& \frac{i}{\pi} \frac{k_f k_i}{k_f^2 - k_i^2}
\frac{1}{k_f + k_i} \left[q_{k_i}(\vert n \vert)-(-1)^n q_{k_f}(\vert n \vert) \right]s(n) \nonumber \\
s(n)&=&\theta(n)-(-1)^n \theta(-n)~~,~~q_k(n)=\frac{1}{g_0^n} (\sqrt{k^2+g_0^2}-k)^n
\label{eq:eq27}
\end{eqnarray}
with $A_{k_f k_i}(0)=0$. The various $n$ contributing to the sum
in  eq.~(\ref{eq:eq25}) correspond to the usual higher Floquet
modes. As it can be seen from eq.~(\ref{eq:eq27}) the amplitude
$A_{k_f k_i}$ for the inelastic continuum-continuum transitions
decays exponentially with $\vert n \vert$ for $\frac{g_0}{k_<} <
1$ (where $k_<=\min(k_f,k_i))$:
\begin{equation}
A_{k_f k_i}(n) {\displaystyle{\stackrel{\approx}{_{_{g_0 \to 0}}}}} \left(\frac{g_0}{k_{<}}\right)^{\vert n \vert}
\label{eq:eq28}
\end{equation}
in accordance with  the fast convergence of the Floquet sum
observed  (but not explained) in the analysis of the dynamics of the oscillating
delta-barrier in the literature. It must be noticed here, that, for $\frac{g_0}{k_<} > 1$, although the approximation (\ref{eq:eq28}) does not hold, the perturbative expansion is still valid. The reason is that, also in this case, the contribution of the diagrams with increasing number of transitions decreases. 
Since the dimensionless
coupling $g_0$ is given by the original coupling multiplied by a
factor proportional to $\omega^{-1/2}$ our treatment is
necessarily non-adiabatic and becomes exact either in the weak
coupling or in the rapid oscillations (very large oscillation
frequency $\omega$) limit. One important issue to be noticed here
is that the expansion of the $S$-matrix in terms of the number of
transitions, as described above, allows the decomposition of the
transmission process into elementary sub-processes giving a
consistent meaning to our approximation procedure. The emerging
perturbative scheme can also be understood in terms of an
expansion in powers of the ratio  $\frac{g_0}{k_i}$ of the coupling $g_0$ over
magnitude of the incoming wave vector $k_i$. As can be directly
confirmed from eqs.~(\ref{eq:eq16}) and (\ref{eq:eq17}), the
c/c transitions are of order $\frac{g_0}{k_i}$ while
c/b or b/c transitions are of order
$\left(\frac{g_0}{k_i}\right)^{3/2}$. The magnitude of the various
diagrammatic contributions to the $S$-matrix is then quantified by
the leading power of $\frac{g_0}{k_i}$, a power that increases as
the number of transitions increases.

In the next order (two transitions) the c/b or b/c transitions are
also possible. Typically, the corresponding
sub-process is demonstrated by the fourth diagram on the right hand
side shown in Fig.~2. The contribution of this term to the
amplitude $S_{fi}$ is given as:
\begin{equation}
S_{fi}^{(2t,0c,2b)} = i \int_{-\infty}^{\infty} d\tau_2 \int_{-\infty}^{\tau_2} d\tau_1 e^{i(\epsilon_f \tau_2 - \epsilon_i \tau_1)}
e^{2 i \theta_{k_f}(\tau_2)} \Phi_{k_f b}(\tau_2) \Delta_b^{(0)}(\tau_2,\tau_1) \Phi_{b k_i}(\tau_1) e^{-2 i \theta_{k_i}(\tau_1)} \theta(k_f)
\label{eq:eq29}
\end{equation}
This amplitude is non-zero only if the time variables are in the
region $(2 \pi \nu, \pi + 2 \pi \nu)$.  The needed bound-state
propagator fulfils the periodic boundary condition:
$\Delta_b^{(0)}(\tau_2,\tau_1)=\Delta_b^{(0)}(\tau_2 +  2 \pi
m,\tau_1)$ ($m=0,\pm 1,..$) and is given by eq.~(\ref{eq:eq22})
where the allowed number of terms which must be summed up, depends
on the difference $\tau_2-\tau_1$. These facts complicate the
calculation of the amplitude (\ref{eq:eq29}). We can considerably
simplify things by replacing the time-dependent bound-state energy,
in the framework of our approximation scheme, by its mean value
over a period:
\begin{equation}
\bar{\epsilon}_b=\frac{1}{2 \pi} \int_0^{2 \pi} d\tau \epsilon_b(\tau)=-\frac{1}{8}g_0^2
\label{eq:eq30}
\end{equation}
This approximation is a first order estimation, being justified in
the case of  very fast or very slow (static limit) oscillations of the potential
which lead to an effective, time-independent, bound-state energy.
Higher order corrections can be obtained by expanding the
bound-state wave function around the effective coupling value
$\frac{g_0}{2}$. Following this scheme the b/c transition amplitude becomes in first order:
\begin{equation}
\bar{\Phi}_{k b}(\tau)=2 i k \sqrt{\frac{g_0}{4 \pi}} \frac{\dot{g}_{\tau}}{k - i g_0/2} \frac{e^{-2 i \theta_k(\tau)}}{k^2 + g_{\tau}^2}
\label{eq:eq31}
\end{equation}
Introducing now the Fourier transformations:
\begin{eqnarray}
e^{2 i \theta_{k_f}(\tau)} \bar{\Phi}_{k_f b}(\tau) &=& \sum_{n=-\infty}^{\infty} B_{k_f b}(n) e^{-i n \tau} \nonumber \\
\bar{\Phi}_{b k_i}(\tau)e^{-2 i \theta_{k_i}(\tau)}  &=& \sum_{n=-\infty}^{\infty} B_{b k_i}(n) e^{-i n \tau}
\label{eq:eq32}
\end{eqnarray}
with:
\begin{eqnarray}
B_{k_f b}(n)&=&\frac{1}{2 \pi} \int_0^{2 \pi} d\tau e^{2 i \theta_{k_f}(\tau)} \bar{\Phi}_{k_f b}(\tau) e^{i n\tau} \nonumber \\
B_{b k_i}(n)&=&\frac{1}{2 \pi} \int_0^{2 \pi} d\tau \bar{\Phi}_{b k_i}(\tau)e^{-2 i \theta_{k_i}(\tau)} e^{i n\tau}
\label{eq:eq33}
\end{eqnarray}
we obtain for the amplitude $B_{k b}(n)$:
\begin{equation}
B_{k b}(n)=i \sqrt{\frac{g_0}{4 \pi}} \frac{1}{k - i g_0/2} q_k(\vert n \vert) \left[ 1 -(-1)^n \right] = B^*_{b k}(-n)
\label{eq:eq34}
\end{equation}
where $q_k(n)$ is given in
eq.~(\ref{eq:eq27}).  It is straightforward to show that also the
b/c transition amplitudes $B_{k b}(n)$ decay
exponentially with the Floquet index $n$ when the applied
perturbation scheme is valid, i.e.:
\begin{equation}
\vert B_{k b}(n) \vert \displaystyle{\stackrel{\approx}{_{_{g_0 \to 0}}}} \frac{1}{\sqrt{k}} \left(\frac{g_0}{k}\right)^{\vert n \vert + \frac{1}{2}}
\label{eq:eq35}
\end{equation}
Using the expressions (\ref{eq:eq32}) we can calculate the
contribution  of the c/b or b/c transitions
corresponding to the third diagram on the right hand side of
Fig.~2 to the $S$-matrix as follows:
\begin{equation}
S_{fi}^{(2t,0c,2b)} = -2\pi i \sum_{n=-\infty}^{\infty} \frac{B_{k_f k_i}(n)}{\vert k_f \vert} \theta(k_f) \delta(k_f - \sqrt{k_i^2 + 2 n}) + (k_f \leftrightarrow -k_f)
\label{eq:eq36}
\end{equation}
where
\begin{equation}
B_{k_f k_i}(n)=\sum_{n_0=-\infty}^{\infty} \frac{B_{k_f b}(n+n_0) B_{b k_i}(-n_0)}{\tilde{\epsilon}_i - n_0 + i \eta}
\label{eq:eq37}
\end{equation}
and $\tilde{\epsilon}_i=\epsilon_i - \bar{\epsilon}_b$. The small
parameter $\eta$ in the denominator of (\ref{eq:eq37}) is
introduced in order to ensure convergence for $\tau_i \to -\infty$
and is equivalent to the demand that the continuum Green's
functions vanish in this limit. In the absence of $\eta$ the
amplitude $B_{k_f k_i}$ possesses a pole at $\tilde{\epsilon}_i =
n_0$. The occurrence of the pole is, from a mathematical point
of view, a result of the perturbative expansion. In fact, the
time-dependence of the potential produces an effective Hamiltonian
containing an infinite series of geometrical phases
\cite{Fujikawa2008}. These terms are the origin of a
non-vanishing imaginary part $\eta^R$ that naturally appears when
higher order terms, involving virtual transitions between the continuum and the bound state (virtual ``multi-photon" exchange), are
summed up. As a consequence the denominator of the amplitude
(\ref{eq:eq37}) never vanishes on the real axis of the energy. At
the same time these higher order terms shift the energy
$\tilde{\epsilon}_i$ by $\delta \epsilon$. This is equivalent with
a change in the effective bound state energy of the form:
\begin{equation}
\bar{\epsilon}_b \rightarrow \bar{\epsilon}_b^R=-\frac{1}{8} g_0^2 - \delta \epsilon
\label{eq:eq38}
\end{equation}
Including the higher order corrections in (\ref{eq:eq37}) we
obtain  a normalized redefinition of the continuum-bound-continuum (c/b/c)
transition:
\begin{equation}
B_{k_f k_i}(n) \rightarrow B_{k_f k_i}^R(n)=\sum_{n_0=-\infty}^{\infty} \frac{B_{k_f b}(n+n_0) B_{b k_i}(-n_0)}{\tilde{\epsilon}^R_i - n_0 + i \eta_n^R} Z_n
\label{eq:eq39}
\end{equation}
where $Z_n$ is a normalization factor and $\tilde{\epsilon}^R_i$,
$\eta^R$  are the corrected values. In Appendix A we shall present
the detailed calculation of the renormalized factors indicating the significance of the virtual ``multi-photon" exchange processes. Here it
suffices to note that in the limit  $\frac{g_0}{k_i} < 1$ we have
$B_{k_f b}(n+n_0) B_{b k_i}(-n_0) \sim O(g_0^{\vert n+n_0 \vert +
\vert n_0 \vert + 1})$ which in turn means that, at least, $B_{k_f
k_i} \sim O(g_0^3)$ (for $n=0$, $n_0=\pm 1$ or $n=n_0=-1$). On the
other hand the behaviour of the corrected factors in the
denominator of (\ref{eq:eq39}) depends only weakly on $n$ and
$n_0$ (see eq.~(\ref{eq:A.4}) and the discussion below this equation in Appendix A): 
$\bar{\epsilon}_b^R \sim O(g_0^2)$, $\eta^R \sim
O(g_0^3)$. As a consequence, the corrected values play an
essential role only when the incoming energy is close to an integer
value.

The expression (\ref{eq:eq39}) is finite and well-behaved for all
the  values of the incoming energy. However when the energy of the
incoming particles differs from the effective bound-state energy
by a positive integer the probability amplitude to arrive
at the final state passing through the bound state has a sharp
maximum. The impact of this maximum on the transmission properties
will be discussed below. To complete the calculation of
the $S$-matrix with two transitions we have to calculate also the contribution of the
second diagram on the right hand side of Fig.~2 containing two
c/c transitions. After some straightforward steps we
obtain:
\begin{equation}
S_{fi}^{(2t,2c,0b)} = -4 \pi i \sum_{n=-\infty}^{\infty} \Gamma_{k_f k_i}(n) \frac{\theta(k_f)}{\vert k_f \vert} \delta(k_f-\sqrt{k_i^2 + 2 n})
\label{eq:eq40}
\end{equation}
with:
\begin{equation}
\Gamma_{k_f k_i}(n)=\int_0^{\infty} dk \sum_{l=-\infty}^{\infty} \frac{A_{k_f k}(n-l) A_{k k_i}(l)}{\epsilon_i-\epsilon_k + l + i 0^+}
\label{eq:eq41}
\end{equation}
and $A_{k k^{\prime}}$ given by (\ref{eq:eq27}). Due to the integration over $k$ the regularizing imaginary part in the denominator of eq.~(\ref{eq:eq41}) can be taken as zero. It can be further
confirmed,  both numerically and analytically (see Appendix B),
that the imaginary part of the above amplitude
$$\Im \Gamma_{k_f k_i}(n) = -\pi \sum_{\stackrel{l=-\infty}{k_i^2 + 2l \geq 0}}^{\infty} \frac{1}{\vert k_l \vert} A_{k_f k_l}(n-l) A_{k_l k_i}(l)~~~~;~~~~k_l=\sqrt{k_i^2 + 2 l}$$
which is controlled by the pole at $\epsilon_k=\epsilon_i + l$
dominates over its real part
$$\Re \Gamma_{k_f k_i}(n) = Pr \int_0^{\infty} dk \sum_{l=-\infty}^{\infty} \frac{A_{k_f k}(n-l) A_{k k_i}(l)}{\epsilon_i-\epsilon_k + l}$$ that avoids the pole ($Pr$ denotes the principal value integration). In the case $\frac{g_0}{k_i} < 1$, the imaginary part is of order $O(g_0^{\vert n - l \vert + \vert l \vert})$ which, for $n=0$, $l=\pm 1$, is of order $O(g_0^2)$, while the real part is, at least, of order $O(g_0^4)$.

Already at this general stage of the analysis one can gain
some  insight into the expected behaviour of the transmission
coefficient. Whenever a propagator is contained in a diagram
contributing to the transmission amplitude we observe the
appearance of denominators which are in fact associated with the
$\frac{1}{\hat{H}_0 - \tilde{E}}$ factors occurring in the
expansion (\ref{eq:eq10}). These denominators are mainly
responsible for resonant structures in the transmission
coefficient. The most important contributions come from diagrams
involving propagators close to the incoming and outgoing states
since in this case a pole structure in the complex energy plane
emerges. If the propagators are bracketed between transitions
which do not involve the incoming or the outgoing state then the
pole structure is integrated out leading to a smoother
contribution to the transmission coefficient.  However, the effects
of these denominators is also controlled by the transition amplitudes
which occur in the numerator and play the role of residues in
the final expression for the contribution of a given diagram. This
interplay between propagation and transition (poles and residues)
determines the overall behaviour of the transmission coefficient.

\section{Transmission properties}

Having calculated the amplitudes of all sub-processes involving
up to two transitions it is straightforward to calculate the associated
$S$-matrix amplitude up to this order. In general for scattering off
a potential varying periodically with time the $S$-matrix amplitude
is given as:
\begin{equation}
S_{fi}=T_{ii}(0) \delta(k_f - k_i) + R_{ii}(0) \delta(k_f + k_i) + \sum_{n \neq 0} \left[T_{fi}(n) \delta(k_f - \sqrt{k_i^2 + 2 n}) + R_{fi}(n)
\delta(k_f + \sqrt{k_i^2 + 2 n}) \right]
\label{eq:eq42}
\end{equation}
where $T_{ii}(0)$, $T_{fi}(n)$ are respectively the elastic and
inelastic transmission amplitudes while $R_{ii}(0)$, $R_{fi}(n)$
are the corresponding reflection amplitudes. They can be expressed
in terms of the calculated sub-processes as follows:
\begin{eqnarray}
T_{ii}(0)&=&1 - \frac{2 \pi i}{\vert k_i \vert} B^R_{k_i k_i}(0) - \frac{4 \pi i}{\vert k_i \vert} \Gamma_{k_i k_i}(0) + ... \nonumber \\
R_{ii}(0)&=&- \frac{2 \pi i}{\vert k_i \vert} B^R_{-k_i k_i}(0) - \frac{4 \pi i}{\vert k_i \vert} \Gamma_{-k_i k_i}(0) + ... \nonumber \\
n \neq 0: \quad T_{fi}(n)&=&\frac{2 \pi i}{\vert k_f \vert} A_{k_f k_i}(n) - \frac{2 \pi i}{\vert k_f \vert} B^R_{k_f k_i}(n) - \frac{4 \pi i}{\vert k_f \vert} \Gamma_{k_f k_i}(n) + ... \nonumber \\
n \neq 0: \quad R_{fi}(n)&=&\frac{2 \pi i}{\vert k_f \vert} A_{-k_f k_i}(n) - \frac{2 \pi i}{\vert k_f \vert} B^R_{-k_f k_i}(n) - \frac{4 \pi i}{\vert k_f \vert} \Gamma_{-k_f k_i}(n) + ...
\label{eq:eq43}
\end{eqnarray}
The total transmission coefficient is obtained as:
$$T_{tot}(\epsilon_i)= \vert T_{ii}(0) \vert^2 + \sum_{\stackrel{n=-\infty}{n \neq 0}}^{\infty} \frac{\vert k_f(n) \vert}{k_i} \vert T_{fi}(n) \vert^2$$

\noindent
{\it Contribution of continuum-continuum versus continuum-bound state (or b/c) transitions} \\
The decomposition (\ref{eq:eq43}) allows us to isolate the
contribution to  the transmission of the sub-processes involving
c/b or b/c transitions and to compare with the
corresponding contribution of sub-processes involving exclusively
c/c transitions. This is not possible in any other
approach. In addition, as in Floquet theory, we can also here
discriminate between elastic and inelastic contributions to the
transmission. Let us first concentrate on the dominating elastic
transmission channel. According to eq.~(\ref{eq:eq43}) the elastic
part of the transmission coefficient is given as:
\begin{eqnarray}
T_{ii}(0)&=&1 - \frac{2 \pi i}{\vert k_i \vert} B^R_{k_i k_i}(0) - \frac{4 \pi i}{\vert k_i \vert} \Gamma_{k_i k_i}(0) + ... \nonumber \\
&=& 1 - \frac{2 \pi i}{\vert k_i \vert} \sum_{n_0} \frac{B_{k_i b}(n_0) B_{b k_i}(-n_0)}{\tilde{\epsilon}^R_i(n_0) - n_0 + i \eta_0^R(n_0)} Z_0 - \frac{4 \pi i}{\vert k_i \vert} \Gamma_{k_i k_i}(0) + ...
\label{eq:eq44}
\end{eqnarray}
The discussion below eqs.~(\ref{eq:eq39}) and (\ref{eq:eq41})
indicates  that the dominant contributions in (\ref{eq:eq44}) are due to the $\Re
B^R_{k_i k_i}(0)$ and $\Im \Gamma_{k_i k_i}(0)$ for all values of the incoming energy
$\epsilon_i$ except for $\epsilon_i \approx n_0 - \vert \bar{\epsilon}_b^R \vert$ where the $\Im B^R_{k_i k_i}(0)$ dominates. Thus it is
suggestive to define the relative weight  $w_0(\epsilon)$ of
contributions to the transmission coefficient involving
transitions between the continuum and the bound state to contributions
involving exclusively transitions between continuum states as
follows:
$$w_0(\epsilon_i)=\frac{\vert 2 \pi \Re B^R_{k_i k_i}(0) \vert}{ \vert k_i + 4 \pi \Im \Gamma_{k_i k_i}(0) \vert}$$
In Fig.~3 we plot for illustration $w_0$ as a function of the
incoming energy $\epsilon_i$ for two different values of the
coupling $g_0$. We see that the processes involving
c/c transitions dominate for the complete energy regime
except of a small region around the integer value $\epsilon_i=1$.
\begin{figure}[htbp]
\centerline{\includegraphics[width=14 cm,height=9
cm]{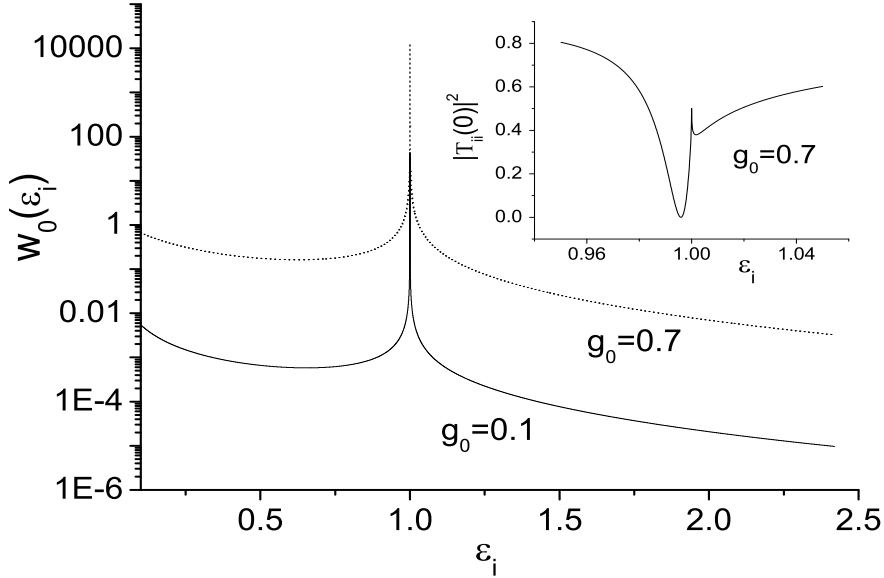}} \caption{The relative weight $w_0$ of the
contribution of sub-processes  involving continuum-bound state
transitions to those involving exclusively continuum-continuum
transitions to the elastic part of the transmission coefficient as
a function of the incoming energy. Two values of the coupling
$g_0$ are used: 0.1 (solid line) and 0.7 (dotted line). The inset
displays the elastic part of the transmission coefficient in an energy 
domain around $\epsilon_i=1$ for $g_0=0.7$. }
\label{fig:3}
\end{figure}

\noindent
{\it Existence of a single transmission zero} \\
The dramatic increase of the amplitude of the sub-processes
involving  transitions between continuum and the bound state
around the integer value $1$ of the incoming energy has important
consequences on the behaviour of the transmission coefficient with varying energy (see the inset in Fig.~3). To clarify this let us analyse the elastic channel in more
detail. In eq.~(\ref{eq:eq44}) the first term corresponds to free
transmission, the second to sub-processes with two transitions involving
one c/b and one b/c
transition, while the third term involves two c/c
transitions. The numerator of the second term is of order
$O(g_0^3)$ while the third term is always of the order of
$O(g_0^2)$ \footnote{To simplify the notation we use the ordering
with respect to powers of $g_0$ although it is always meant the
ordering with respect to powers of $\frac{g_0}{k}$. }. The most
interesting behaviour is associated with the denominator of the
second term. As already discussed, the imaginary part $\eta_0^R$ is
of order $O(g_0^3)$ (since it does not occur for sub-processes up
to $O(g_0^2)$). Therefore whenever $\tilde{\epsilon}^R_i=n_0$ the
second term becomes of order one ($O(g_0^0)$) and dominates. In
this case the exact value of $\eta_0^R$ is crucial since the
second term becomes proportional to $\frac{1}{\eta_0^R}$. Within
our approach it can be determined by going to higher order terms
containing more than two transitions involving the bound state. Since
the number of transitions between continuum and bound state is
always even, the next to leading order contains necessarily four
such transitions. The calculation of the contribution to the $S$-matrix
amplitude of diagrams involving four transitions between continuum and
the bound state is straightforward but lengthy and is not
presented here. The final result up to the order $O(g_0^4)$ is
obtained through the summation of terms originating from diagrams
with four and six transitions in total as illustrated in Fig.~4
\footnote{A careful counting shows that the contribution of
diagrams involving 5 transitions to the finite value of $\eta_0^R$ is of
higher order.}.
\vspace*{0.5cm}
\begin{figure}[htbp]
\centerline{\includegraphics[width=14 cm,height=2
cm]{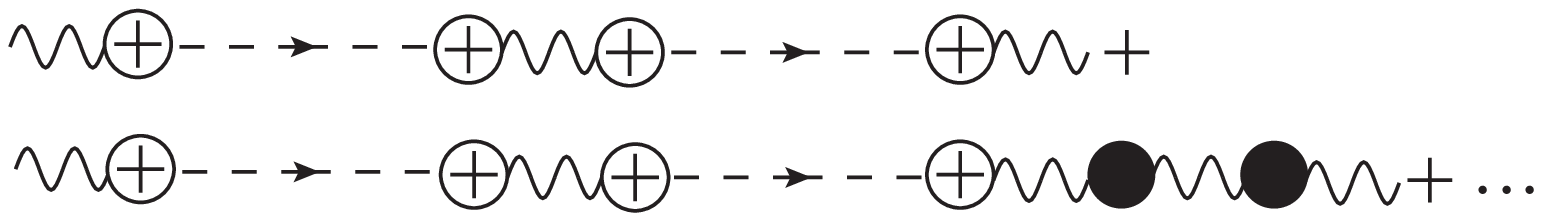}} \caption{The diagrams contributing up to the
order $O(g_0^4)$ to $\eta_0^R$ and to the  shift in
$\tilde{\epsilon}^R_i$. The dots indicate that a sum over all
allowed permutations in the 6-transition diagrams must be performed.}
\label{fig:4}
\end{figure}
The interested reader can find a short description of the
calculation in  Appendix A. 

In the limit $\frac{g_0}{k_i}< 1$, it is
straightforward to study the  behaviour of the transmission
coefficient (\ref{eq:eq44}) in the case $\tilde{\epsilon}^R_i=n_0$ which
%One has to substitute eqs.~(\ref{eq:eq45}) in (\ref{eq:eq44}) and
%expand in powers of $g_0$ neglecting terms of order $O(g_0^4)$.
also depends on the value of $n_0$. Therefore let us first
investigate the case $n_0=1$ for effective incoming energies $0 <
\tilde{\epsilon}^R_i  \leq 1$. When $1 - \tilde{\epsilon}^R_i = O(g_0)
\gg \eta_0^R(1)$ both the imaginary part of the denominator as
well as the normalization factor in the second term of
eq.~(\ref{eq:eq44}) can be neglected and the elastic part of the
transmission is well described by the formula:
\begin{equation}
T_{ii}(0) \approx 1 - \frac{2 \pi i}{k_i} \sum_{n_0=\pm 1} \frac{B_{k_i b}(n_0) B_{b k_i}(-n_0)}{\epsilon_i - n_0} + \frac{4 \pi}{k_i} \Im \Gamma_{k_i k_i}(0)
\label{eq:eq45}
\end{equation}
When $1 - \tilde{\epsilon}^R_i = O(g_0^2)$ the
contribution of the real part of the second term increases and
being negative leads to a decrease of the transmission amplitude.
The effect is maximized when $1 - \tilde{\epsilon}^R_i$ becomes of
the order of $O(g_0^4)$. Then the denominator of the second term
is controlled, almost exclusively, by $\eta_0^R$, which is of the
same order of magnitude with the numerator
$B_{k_i b}(1) B_{b k_i}(-1) \approx O(g_0^3)$. Thus, in a narrow
energy region around $\epsilon_i \approx 1$ the term describing
transitions between continuum and bound state dominates (in the
considered order of perturbation theory) driving the elastic
transmission to zero: $T_{ii}(0) \approx 0$ (for more details see
Appendix A). Following a similar reasoning one can show that also
the inelastic transmission as well as reflection amplitude become
zero in the considered case while the elastic reflection $\vert
R_{ii}(0) \vert \approx 1$. 

When the incoming energy increases beyond the position of the
transmission  zero the expression ({\ref{eq:eq45}) for the elastic
transmission amplitude is again valid. However, the value of the
transmission coefficient will slightly decrease due to the fact
that the imaginary part of $\Gamma_{k_i k_i}(0)$ increases with
increasing energy and contributes negatively to the transmission
amplitude. When the incoming energy is near an integer value $n_0
> 1$, the transmission profile considerably differs from the one
around $n_0=1$. This is due to the following facts. Firstly the
coefficient $B_{k_i b}(n_0) B_{b k_i}(-n_0)$ is exponentially
suppressed when $n_0$ increases and, consequently, the numerator
of the second term in eq.~(\ref{eq:eq44}) becomes of higher and
higher order in $g_0$. Secondly, $\eta_0^R(n_0)$ is always of
order $O(g_0^3)$ almost independently from the definite value of
$n_0$. Therefore, the second term is
always suppressed for $n_0 > 1$ with respect to the other two
terms. This is an issue which has not been resolved in the previous
literature. In \cite{Martinez2001} it is mentioned that the
residues of the higher resonances considered as poles of the
transmission amplitude in the complex energy plane decrease with
increasing energy, however no specific explanation has been provided.

\noindent
{\it Location of the transmission zero} \\
As already discussed, due to the higher order corrections, the transmission zero does not occur for $\epsilon_i=1$ but it is slightly shifted to lower values of the incoming energy. In fact, the exact location of the transmission zero is determined by the effective energy $\tilde{\epsilon}^R_i$. Within our treatment it is straightforward to obtain not only $\eta_0^R$ but also the shift in the effective energy $\tilde{\epsilon}^R_i$ as well as the normalization factor $Z_0$ appearing in eq.~(\ref{eq:eq44}): 
\begin{eqnarray}
\eta_0^R(n_0)& \approx & \beta(n_0) (1 + \gamma_0(n_0)) \nonumber \\
\tilde{\epsilon}^R_i(n_0)& \approx &\epsilon_i + \frac{1}{8}g_0^2 + \alpha(n_0) \nonumber \\
Z_0 & \approx & 1 - \frac{2 \pi i}{k_i} \left[ 4 i \Im \Gamma_{k_i k_i}(0) - \sum_{l \neq n_0} \frac{B_{k_i b}(-l) B_{b k_i}(l)}{\tilde{\epsilon}^R_i(n_0)- l + i \eta_0^R(n_0)}
\right]
\label{eq:eq46}
\end{eqnarray}
with:
\begin{eqnarray}
\beta(n_0)&=&\sum_l \frac{2 \pi}{\sqrt{k_i^2 + 2 l}} B_{k_i b}(n_0+l) B_{b k_i}(-n_0-l) \nonumber \\
\gamma_0(n_0)&=&\frac{4 \pi}{k_i} \Im \Gamma_{k_i k_i}(0) \nonumber \\
\alpha(n_0)&=&2 \sum_l Pr \int_0^{\infty} dk \frac{B_{k b}(n_0+l) B_{b k}(-l-n_0)}{\epsilon_k - \epsilon_i - l}
\label{eq:eq47}
\end{eqnarray}
More details of the calculation of these quantities are given in Appendix A. It is important to notice here that the
effect of the higher order corrections leading to the shift of the effective incoming energy and to the appearance of the
imaginary part $\eta_0^R$ in the denominator of the second term in eq.~(\ref{eq:eq44}) starts to show up in the immediate neighbourhood of the transmission zero, i.e. when $1 - \tilde{\epsilon}^R_i = O(g_0^2)$. In the literature the transmission zero is associated with a Fano resonance \cite{Bagwell1992}. Its
exact position however has not been obtained from first principles. In \cite{Martinez2001} it is estimated by a
continued fractions approach. Within our treatment it is in principle possible (see eq.~(\ref{eq:eq46}) and Appendix A) to
obtain the position of the transmission zero to arbitrary accuracy using increasingly order terms in the proposed perturbation scheme. In addition our treatment provides a physical picture for the
origin of this shift: it is attributed to a hierarchical sequence
of sub-processes involving repetitive transitions of the incoming
particle from the continuum to the bound state and back. These can be identified with a virtual ``multi-photon" exchange of the propagating particle with the driving potential. Sub-processes
with increasing number of transitions become more and more
suppressed.\\

\noindent
{\it Further features of the transmission coefficient}\\
In addition to the above analysed transmission zero there occur pronounced
local  peaks in the transmission coefficient with increasing incoming energy. A typical structure is shown in
Fig.~5(a) where a zoom in the region around the zero of the
transmission coefficient in the elastic channel is performed. This
structure is attributed to the competition between the real part of
the second term in eq.~(\ref{eq:eq44}) and the contribution
proportional to $\Im \Gamma_{k_i k_i}(0)$ contained in the third
term. The later possesses a cusp at $\epsilon_i=1$ (see Appendix B
and Fig.~5(b)) and contributes destructively (negative sign) in
the transmission amplitude while the former decreases rapidly as
the incoming energy increases beyond the transmission zero value.
Depending on the position of the transmission zero with respect to
the value $1$ where the discontinuity of $\Im \Gamma_{k_i k_i}(0)$
is located, this competition may lead to a pronounced effect (the
observed peak structure) or not. For the occurrence of the peak
structure the distance between transmission zero and
$\epsilon_i=1$ should be at least of the order of $g_0^2$. It must
be noticed that the discontinuity in $\Im \Gamma_{k_i k_i}(0)$ is
a consequence of integrated poles in the corresponding $S$-matrix
amplitudes.
\begin{figure}[htbp]
\centerline{\includegraphics[width=12 cm,height=7
cm]{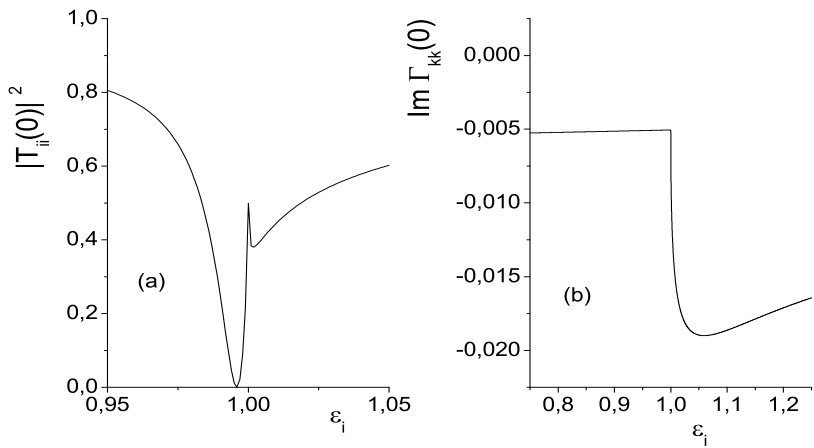}} \caption{(a) The elastic part of the
transmission coefficient as a  function of the incoming energy for
$g_0=0.7$ in the region around the transmission zero and (b) 
$\Im \Gamma_{k_i k_i}(0)$ as a function of the incoming energy in
the same region. } \label{fig:5}
\end{figure}

\section{Concluding remarks}
In the present work we have developed a perturbative scheme for
the  calculation of the $S$-matrix for a scattering problem involving
a time-dependent external potential. Our approach is inspired by
the treatment of scattering in quantum field theory and contains as a key ingredient
a temporary base description of the 
transition amplitudes. The order of each term in the perturbative expansion is determined by the number of transitions between different states of the temporary
static spectrum. We have applied this approach to the case of a delta
function with harmonically oscillating strength. Here the
temporary spectrum may contain also a single bound state. An
interesting feature of the developed perturbation method is that
the transitions between a continuum state of the temporary
spectrum and the bound state depend on a different power of the
coupling of the potential than the transitions between two
continuum states. Therefore the perturbative approach has to be
applied with special care in order to avoid artefacts. The proposed scheme has several advantages
providing a tool for a detailed study of the scattering process
based on the intuitive particle picture instead of the usual wave
picture which is implied by the Floquet method or the direct integration
of the Schr\"{o}dinger equation. These advantages are here
demonstrated by establishing a systematic procedure for the exact
calculation of the transmission zero associated with the appearance
of a Fano resonance in the considered system. In addition, within
the proposed scheme, it is possible to show that the transition
amplitudes of the inelastic channels decrease exponentially with
increasing order in the Floquet expansion (real ``multi-photon" exchange)
a fact that explains the appearance of only one resonant energy in
the transmission coefficient. Furthermore our approach reveals the
significance of virtual ``multi-photon" exchange processes which are responsible for 
a slight shift of the location of the transmission zero. 
Finally the developed method enables  also the detailed analysis
of local peak structures occurring in the dependence of the
transmission coefficient on the incoming energy which cannot be
attributed to resonances (poles of the transmission amplitudes)
and are usually not investigated in the literature. A disadvantage
of the proposed method is that it is technically more demanding
than the other existing methods mentioned above possessing several
anomalies similar to those of quantum field perturbation theory.
However these technical complications are balanced by the deeper
insight gained for the considered physical problem opening the
perspective of a systematic decomposition of the time-dependent
scattering process in fundamental sub-processes building up the
skeleton of transport in driven quantum systems.

\noindent {\bf Acknowledegments} This research has been 
co-financed by the  European Union (European Social Fund – ESF)
and Greek national funds through the Operational Program
"Education and Lifelong Learning" of the National Strategic
Reference Framework (NSRF) - Research Funding Program: Heracleitus
II. Investing in knowledge society through the European Social
Fund. Further financial support by the Greek Scholarship
Foundation IKY in the framework of an exchange program with
Germany (IKYDA) is also acknowledged.

\appendix
\section{}
In this Appendix we shall briefly present  the calculations that
lead to the corrected values appearing in eq.~(\ref{eq:eq39}) of the
text. To begin with,  we note that the need to consider higher
order corrections  can be understood in terms of the following
schematic expansion:
\begin{equation}
\label{eq:A.1}
\frac{{{B^2}}}
{{\left( {{{\tilde \varepsilon }_i} + \delta \varepsilon } \right)
- {n_0} + i\left( {\eta  + \delta \eta } \right)}} \approx
\frac{{{B^2}}} {{{{\tilde \varepsilon }_i} - {n_0} + i\eta }} -
\frac{{{B^2}\delta \varepsilon }} {{{{\left( {{{\tilde \varepsilon
}_i} - {n_0} + i\eta } \right)}^2}}} - i\frac{{{B^2}\delta \eta }}
{{{{\left( {{{\tilde \varepsilon }_i} - {n_0} + i\eta }
\right)}^2}}} + ...
\end{equation}
It is obvious that when ${\tilde \varepsilon _\iota } \approx {n_0}$, the corrections
$\delta \varepsilon $ and $\delta \eta $ must be taken into
account.  The first process that contains a double pole term involves in total 4 transitions, all
of b/c or c/b type:

\begin{eqnarray}
\label{eq:A.2}
\displaystyle S_{fi}^{(4t,0c,4b)} & \approx & - 4\pi i\sum_{n,{n_1},{n_2},{n_3}} {\frac{{{B_{{k_f}b}}\left( {n - {n_3}} \right)}}
{{{{\tilde \varepsilon }_i} + {n_3} + i\eta }}} \left[
{\int_0^\infty  {dk\frac{{{B_{bk}}\left( {{n_3} - {n_2}}
\right){B_{kb}}\left( {{n_2} - {n_1}} \right)}} {{{\varepsilon _i}
- {\varepsilon _k} + {n_2} + i\eta }}} }
\right]\frac{{{B_{b{k_i}}}\left( {{n_1}} \right)}} {{{{\tilde
\varepsilon }_i} + {n_1} + i\eta}} \nonumber \\  
 & \stackrel{ \displaystyle \approx }{_{{n_1} = {n_3} \to  - {n_0}}} & - 2\pi i\sum_{n,{n_0}} {\frac{{{B_{{k_f}b}}\left( {n + {n_0}} \right){B_{b{k_i}}}\left( { - {n_0}} \right)}}
{{{{\left( {{{\tilde \varepsilon }_i} - {n_0} + i\eta }
\right)}^2}}}\left[ { - \alpha \left( {{n_0}} \right) - i\beta\left({{n_0}}\right)} \right]} 
\end{eqnarray}
                             
\noindent
In the last expression we wrote: 
\begin{equation}
 \label{eq:A.3}
\alpha \left(
{{n_0},{\varepsilon _i}} \right) \equiv 2\sum\limits_l {\Pr
\int\limits_0^\infty  {dk\frac{{{B_{kb}}\left( {{n_0} + l}
\right){B_{bk}}\left( { - l - {n_0}} \right)}} {{{\varepsilon _k}
- {\varepsilon _i} - l}}} } 
\end{equation}
 and 
\begin{equation}
 \label{eq:A.4}
\beta \left( {{n_0},{\varepsilon _i}} \right) \equiv
\sum\limits_l {\frac{{2\pi }} {{{k_l}}}{B_{{k_l}b}}\left( {{n_0} +
l} \right)} {B_{b{k_l}}}\left( { - l - {n_0}} \right),{\text{
}}\left( {{k_l} = \sqrt {k_i^2 + 2l} } \right)
\end{equation}

\noindent
Comparing eq.~(\ref{eq:A.2}) with the expansion (\ref{eq:A.1}) we get the corrected quantities appearing in the text:
\begin{equation}
\label{eq:A.5}
\tilde \varepsilon _i^R\left( {{n_0}} \right) \approx {\varepsilon _i} + \frac{1}{8}{g^2} + \alpha \left( {{n_0},{\varepsilon _i}} \right)
\end{equation}

\begin{equation}
\label{eq:A.6}
\eta _{{n_0}}^R \approx \beta \left( {{n_0},{\varepsilon _i}} \right)\    
\end{equation}

\noindent
Some comments are in order at this point. The function $\alpha $
corrects the pole position:

\begin{equation}
\label{eq:A.7}
{\varepsilon _i} = {n_0} - \frac{1}{8}{g^2} - \alpha \left(
{{n_0},{\varepsilon _i}} \right) 
\end{equation}

\noindent
Its leading behaviour in the limit ${g_0}/{k_i} < 1$ is obtained
for ${n_0} + l =  \pm 1$ and it is almost independent of  ${n_0}$, having the form:

\begin{equation}
\label{eq:A.8}
\alpha  \approx  - {g^2}{\alpha _1} - {g^4}{\alpha _2},{\text{
}}{\alpha _{1,2}} > 0 
\end{equation}
The non-vanishing value of $\beta $ (see eq.~(\ref{eq:A.4})) drives to finite  values  all the terms in the
perturbative calculation of the $S$-matrix. Similarly to $\alpha$, its leading behaviour, being of order $O\left( {g_0^3} \right)$, is obtained for the combination ${n_0} + l = \pm 1$ and it is almost independent
of ${n_0}$.  This behaviour is responsible for the zero of the elastic
transmission amplitude: for ${\varepsilon _i} \approx 1 -
\frac{1}{8}{g^2} - \alpha $, the amplitude
$\frac{2 \pi i}{k_i}B_{{k_i}{k_i}}^R\left(0 \right)$  dominates and being  real,
positive and of order $O(1)$ cancels  the free transmission
contribution driving the total amplitude to zero.  For
${\varepsilon _i} \approx {n_0} > 1$ the phenomenon is suppressed
since $B_{{k_i}{k_i}}^R\left( 0 \right) \sim O\left( {g_0^{2\left(
{{n_0} - 1} \right)}} \right)$. 

For the quantitative analysis of
the scattering process, in the energy region in which the ``pole"
behaviour is visible,  one needs  a more exact value of
$\beta $ and this calls for corrections  coming from higher
order processes. The 5-transitions amplitude that contains 4 b/c or c/b
transitions   and one c/c
transition, gives a higher order correction of the pole
position and we shall not discuss it here. On the other hand the
6-transitions amplitude with 4 b/c or c/b and two c/c transitions contributes
to $\eta$, that is to the imaginary part of
$B_{{k_f}{k_i}}^R\left( n \right)$:

\begin{equation}
\label{eq:A.9}
S_{fi}^{(6t,2c,4b)}  \approx  - 2\pi
i\sum\limits_{{n_0}} {\frac{{{B_{{k_i}b}}\left( {n + {n_0}}
\right){B_{b{k_i}}}\left( { - {n_0}} \right)}} {{{{\left(
{{{\tilde \varepsilon }_i} - {n_0} + i\eta } \right)}^2}}}\left[ {
- i{\gamma _n}\left( {{n_0}} \right)\beta \left( {{n_0}} \right)}
\right]}
\end{equation}
where

\begin{equation}
\label{eq:A.10}
{\gamma _n}\left( {{n_0}} \right)
\equiv \frac{{2\pi }} {{{k_i}}}\Im \left[ {{\Gamma
_{{k_i}{k_i}}}\left( 0 \right) + \frac{{{B_{{k_i}b}}\left( {{n_0}}
\right)}} {{{B_{{k_f}b}}\left( {n + {n_0}} \right)}}{\Gamma
_{{k_f}{k_i}}}\left( n \right)} \right]
\end{equation}

\noindent                     
Combining the above results with eq.~(\ref{eq:A.2}) we find : 

\begin{equation}
\label{eq:A.11}
\eta _n^R
\approx \beta \left( {{n_0}} \right)\left[ {1 + {\gamma _n}\left({{n_0}} \right)} \right]
\end{equation}

For the elastic channel $n = 0$ the last result is apparent in eq.~(\ref{eq:eq46}). When the incoming energy is at the realm of
the ``pole", higher order corrections of the pole residue, coming from first order poles
in the 3-transitions amplitude with two b/c or c/b and one c/c
transitions and in the  4-transitions amplitude with 2 b/c or c/b and 2 c/c
transitions, must be taken into account:

\begin{equation}
\label{eq:A.12}
S_{fi}^{\left( {3t,1c,2b} \right)} + S_{fi}^{(4t,2c,2b)} 
\approx - 2\pi i\sum\limits_{{n_0}} {\frac{{{B_{{k_f}b}}\left( {n + {n_0}} \right){B_{b{k_i}}}( - {n_0})}}
{{{{\tilde \varepsilon }_i} - {n_0} + i\eta }}} \left( {{Z_n} - 1}
\right)
\end{equation}
                                                                                                                                              
where

\begin{eqnarray}
\label{eq:A.13} 
 {Z_n} \approx 1 - \frac{{2\pi i}}
{{{k_i}}}\left[ {2{\Gamma _{{k_i}{k_i}}}(0) +
\frac{{{B_{{k_i}b}}\left( {{n_0}} \right)}}
{{{B_{{k_f}b}}\left( {n + {n_0}} \right)}}\left( {2{\Gamma _{{k_f}{k_i}}}\left( n \right) - {A_{{k_f}{k_i}}}\left( n \right)} \right) - } \right.  \nonumber \\
  \left. {{\text{                }} - \sum\limits_{l \ne  - {n_0}} {\frac{{{B_{{k_f}b}}\left( {n - l} \right){B_{b{k_i}}}\left( l \right)}}
{{{{\tilde \varepsilon }_i} + l + i\eta
}}\frac{{{B_{{k_i}b}}\left( {{n_0}} \right)}}
{{{B_{{k_f}b}}\left( {n + {n_0}} \right)}}} } \right]
\end{eqnarray}

\noindent
Combining the results of eqs.~(\ref{eq:A.5}), ~(\ref{eq:A.11}) and ~(\ref{eq:A.13}) we find the
normalised contribution of the b/c or c/b transitions to the scattering
process as given in eq.~(\ref{eq:eq39}):

\begin{equation}
\label{eq:A.14}
B_{{k_f}{k_i}}^R\left( n \right) = \sum\limits_{{n_0} =  - \infty
}^\infty  {\frac{{{B_{{k_f}b}}\left( {n + {n_0}}
\right){B_{b{k_i}}}\left( { - {n_0}} \right)}} {{\varepsilon
_i^R\left( {{n_0}} \right) - {n_0} + i\eta _n^R\left( {{n_0}}
\right)}}} {Z_n}                                
\end{equation}

\noindent
The above formulas capture the non-trivial
behaviour of the scattering amplitude in the vicinity of a ``pole". For example, the  zero of the elastic transmission amplitude in the region $\left| {1 - \varepsilon _i^R} \right| \ll \eta _0^R$ is easily verified:

\begin{eqnarray}
\label{eq:A.15}
  {T_{ii}}(0) \approx 1 - \frac{{2\pi i}}
{{{k_i}}}\sum\limits_{l \ne 1} {\frac{{{B_{{k_i}b}}\left( l
\right){B_{b{k_i}}}\left( { - l} \right)}} {{{\varepsilon _i} -
l}}}  - \frac{{2\pi i}} {{{k_i}}}\frac{{{B_{{k_i}b}}\left( 1
\right){B_{b{k_i}}}\left( { - 1} \right)}} {{i{\eta ^R}\left( 1
\right)}}{Z_0} + \frac{{4\pi }}
{{{k_i}}} \Im{\Gamma _{{k_i}{k_i}}}(0) \approx \nonumber \\ 
  {\text{       }} \approx 1 - \frac{{2\pi i}}
{{{k_i}}}\sum\limits_{l \ne 1} {\frac{{{B_{{k_i}b}}\left( l
\right){B_{b{k_i}}}\left( { - l} \right)}} {{{\varepsilon _i} -
l}}}  - \frac{{2\pi i}} {{{k_i}}}\frac{{{B_{{k_i}b}}\left( 1
\right){B_{b{k_i}}}\left( { - 1} \right)}} {{\frac{{2\pi i}}
{{{k_i}}}{B_{{k_i}b}}\left( 1 \right){B_{b{k_i}}}\left( { - 1}
\right)\left[ {1 + \frac{{4\pi }}
{{{k_i}}} \Im{\Gamma _{{k_i}{k_i}}}\left( 0 \right)} \right]}} \times  \nonumber  \\
  {\text{         }} \times \left[ {1 + \frac{{8\pi }}
{{{k_i}}} \Im{\Gamma _{{k_i}{k_i}}}\left( 0 \right)
- \frac{{2\pi i}} {{{k_i}}}\sum\limits_{l \ne 1}
{\frac{{{B_{{k_i}b}}\left( l \right){B_{b{k_i}}}\left( { - l}
\right)}} {{{\varepsilon _i} - l}}} } \right] + \frac{{4\pi }}
{{{k_i}}}\Im{\Gamma _{{k_i}{k_i}}}(0) \approx 0 \phantom{aaaaa} 
\end{eqnarray}

\noindent
It is not difficult to show that in the same energy region
and at the same order the inelastic amplitudes also vanish:

\begin{equation}
\label{eq:A.16}
{T_{fi}}\left( n \right) = {R_{fi}}\left( n \right) \approx  - \frac{{{k_i}}}
{{\left| {{k_f}} \right|}}\frac{{{B_{{k_f}b}}\left( {n + 1}
\right)}} {{{B_{{k_i}b}}\left( 1 \right)}}\left[ {1 + \frac{{2\pi
}} {{{k_i}}} \Im{\Gamma _{{k_i}{k_i}}}\left( 0
\right)} \right]              
\end{equation}

\noindent
The term $n = 1$ does not contribute since $B(2) = 0$. For $n > 1$ the coefficients 
\begin{equation}
\label{eq:A.17}
R\left( n \right) = T\left( n \right) \equiv \frac{{\left| {{k_f}} \right|}}
{{{k_i}}}{\left| {{T_{fi}}\left( n \right)} \right|^2}
\end{equation}
are negligible in comparison to $R(0) \approx 1$.

%\appendix 
\section{}
In Appendix A (and in the main text) it is extensively used that the imaginary part of the amplitude 

\begin{equation}
\label{eq:B.1}
{\Gamma
_{{k_i}{k_i}}}\left( 0 \right) = \int\limits_0^\infty
{dk\sum\limits_{l =  - \infty }^\infty  {\frac{{{A_{{k_i}k}}\left(
{ - l} \right){A_{k{k_i}}}\left( l \right)}} {{{\varepsilon _i} -
{\varepsilon _k} + l + i\eta }}} } 
\end{equation}
dominates over its real
part. Besides the numerical confirmation, we present in this
Appendix a direct proof of this statement.  Using the expression
(\ref{eq:eq27}) it is straightforward  to obtain:

\begin{equation}
\label{eq:B.2}
\displaystyle \Im{\Gamma _{{k_i}{k_i}}}(0) =  - \frac{{k_i^2}}{{4\pi }}\sum_{\substack{l \neq 0 \\ k_l^2 = k_i^2 + 2l \geq 0}}
 {\frac{{{k_l}}}{{{l^2}}}} \frac{1} {{{{\left( {{k_i} + {k_l}}
\right)}^2}}}{\left| {{q_{{{\bar k}_l}}}\left( {\left| l \right|}
\right) - {{( - 1)}^l}{q_{{{\bar k}_i}}}\left( {\left| l \right|}
\right)} \right|^2}
\end{equation}
with:
\begin{equation}
\label{eq:B.3}
{q_{\bar k}}\left( {\left| l \right|} \right) = {\left( {\sqrt {{{\bar k}^2} + 1}  - \bar k} \right)^{\left| l \right|}}{\text{  ;   }}\bar k = k/{g_0})
\end{equation}

\noindent
Observing that:
${q_{{{\bar k}_l}}}\left( {\left| l \right|} \right) \sim {\left( {{g_0}/{k_l}} \right)^{\left| l \right|}}$
we immediately find: 
$\Im{\Gamma _{{k_i}{k_i}}}\left( 0 \right) \sim
{\left( {\frac{{{g_0}}}{{{k_i}}}} \right)^2}$. This estimation
is valid if $k_l^2 = k_i^2 + 2l > k_i^2$, i.e. if $l > 0$.
However, negative values of $l$ are permitted if the incoming
energy is large enough. Thus, every time the incoming
energy increases by an integer value, the possibility of an additional negative
value of $l$ occurs. For example, if $k_i^2/2 = 1 + O\left(
{g_0^2} \right)$ the value $l =  - 1$
is permitted and, in fact, it gives the dominant contribution to
the sum (\ref{eq:B.2}) since ${k_l} \approx O\left( {{g_0}} \right)$ and
$\Im{\Gamma _{{k_i}{k_i}}}\left( 0 \right) \approx
O\left( {{g_0}} \right)$. This is the technical explanation for
the cusp appearing in Fig.~\ref{fig:5} of the text. The real part of the
amplitude can also be calculated analytically. Using the
abbreviations
${\lambda _{\bar k}} = \sqrt {{{\bar k}^2} + 1}  - \bar k$, ${\text{ }}{\kappa _{\bar k}} = \sqrt {{{\bar k}^2} + 1}$,
$\displaystyle \sum_{l'} {(...)}  = \sum_{\substack{ {\text{      }}l \ne 0 \\
  k_l^2 = k_i^2 + 2l \geq 0}}  {\left( {...} \right)} $   we find that:
\begin{equation}
\label{eq:B.4}
\Re{\Gamma _{{k_i}{k_i}}}(0) = {I_1} + {I_2}
\end{equation}
The first term on the r.h.s. of the last equation is:

\begin{equation}
\label{eq:B.5}
{I_1} = \frac{{k_i^2}} {{16{\pi ^2}}}\sum\limits_{l'}
{\frac{{k_l^2\ln k_l^2}} {{{l^2}}}\frac{{\left( {{A_1}{B_1}}
\right)}} {{g_0^3}}} 
\end{equation}                                                         
where 
\begin{equation}
\label{eq:B.6}
{A_1} = {q_{{{\bar k}_l}}}\left( {\left| l \right|}
\right)\left[ {\frac{1} {{{{\bar k}_l}\left( {{{\bar k}_l} -
{{\bar k}_i}} \right)}} + \frac{{{{( - 1)}^l}}} {{{{\bar
k}_l}\left( {{{\bar k}_l} + {{\bar k}_i}} \right)}}} \right] -
\frac{2} {{\bar k_l^2 - \bar k_i^2}}{q_{{{\bar k}_i}}}\left(
{\left| l \right|} \right)
\end{equation}
 and 
\begin{equation}
\label{eq:B.7}
{B_1} = {q_{{{\bar
k}_l}}}\left( {\left| l \right| - 1} \right)\left( {\lambda
_{{{\bar k}_l}}^2 - 1} \right)\left[ {\frac{1} {{{{\bar
k}_l}\left( {{{\bar k}_l} - {{\bar k}_i}} \right)}} + \frac{{{{( -
1)}^{l - 1}}}} {{{{\bar k}_l}\left( {{{\bar k}_l} + {{\bar k}_i}}
\right)}}} \right] - \frac{2} {{\bar k_l^2 - \bar
k_i^2}}{q_{{{\bar k}_i}}}\left( {\left| l \right| - 1}
\right)\left( {\lambda _{{{\bar k}_i}}^2 - 1} \right)
\end{equation}

\noindent
The second term has the form
\begin{equation}
\label{eq:B.8}
{I_2} =  - \frac{{k_i^4\ln k_i^2}}
{{4{\pi ^2}}}\sum\limits_{l'} {\frac{1} {{{l^2}}}\frac{{\left(
{{A_2}{B_2}} \right)}} {{g_0^3}}}
\end{equation}
with
\begin{equation}
\label{eq:B.9}
{A_2} = \frac{1}
{2}\frac{{{q_{{{\bar k}_i}}}\left( {\left| l \right| + 1}
\right)}} {{\kappa _{{{\bar k}_i}}^2{{\bar k}_i}}} - \frac{{\left|
l \right| + 1}} {2}\frac{{{q_{{{\bar k}_i}}}\left( {\left| l
\right|} \right)}} {{{\kappa _{{{\bar k}_i}}}{{\bar k}_i}}} -
\frac{{\lambda _{{{\bar k}_i}}^{\left| l \right| + 1}}} {{\kappa
_{{{\bar k}_i}}^2{{\bar k}_i}}} - \frac{1} {4}\frac{{{q_{{{\bar
k}_i}}}\left( {\left| l \right|} \right)}} {{\kappa _{{{\bar
k}_i}}^2{{\bar k}_i}}}\left( {{\lambda _{{{\bar k}_i}}} - {{\bar
k}_i}} \right) + \frac{{{{( - 1)}^l}}} {4}\frac{{{q_{{{\bar
k}_i}}}\left( {\left| l \right|} \right)}} {{\bar k_i^2}}
\end{equation}
and
\begin{eqnarray}
\label{eq:B.10}
{B_2} = \frac{{{q_{{{\bar k}_i}}}\left( {\left| l \right|} \right)}}
{{\kappa _{{{\bar k}_i}}^2{{\bar k}_i}}}\left( {\lambda _{{{\bar
k}_i}}^2 - 1} \right) - \left| l \right|\frac{{{q_{{{\bar
k}_i}}}\left( {\left| l \right| - 1} \right)}} {{{\kappa _{{{\bar
k}_i}}}{{\bar k}_i}}}\left( {\lambda _{{{\bar k}_i}}^2 - 1}
\right) - 4\frac{{\lambda _{{{\bar k}_i}}^{\left| l \right| + 2}}}
{{\kappa _{{{\bar k}_i}}^2{{\bar k}_i}}} - \nonumber  \\
  {\text{    }} - \frac{1}
{2}\frac{{{q_{{{\bar k}_i}}}\left( {\left| l \right| - 1}
\right)}} {{\kappa _{{{\bar k}_i}}^2{{\bar k}_i}}}\left( {\lambda
_{{{\bar k}_i}}^2 - 1} \right)\left( {{\lambda _{{{\bar k}_i}}} -
{{\bar k}_i}} \right) + \frac{{{{( - 1)}^{l - 1}}}}
{2}\frac{{{q_{{{\bar k}_i}}}\left( {\left| l \right| - 1}
\right)}}
{{\bar k_i^2}}\left( {\lambda _{{{\bar k}_i}}^2 - 1} \right)
\end{eqnarray}

\noindent
Observing that
${\lambda _{\bar k}} \sim {g_0},{\text{  }}{\kappa _{\bar k}} \sim 1/{g_0}$
we immediately show that ${A_1}\mathop  \sim \limits_{l = 1}
g_0^3,{B_1}\mathop  \sim \limits_{l = 1} g_0^4$ ${A_2}\mathop
\sim \limits_{l = 1} g_0^3,{B_2}\mathop  \sim \limits_{l = 1}
g_0^4$ and consequently that $\operatorname{\Re} {\Gamma
_{{k_i}{k_i}}}(0) \sim g_0^4$.

\noindent
The case $k_i^2 + 2 l < 0$ can be treated by using the above formulas and making the change  $k_l \to -i \vert k_l \vert$ in the expressions for $\Re \Gamma_{k_i k_i}(0) + 
i \Im \Gamma_{k_i k_i}(0)$. The final result is the same. 


\begin{thebibliography}{99}
\bibitem{delCampo2009} A. delCampo, G. Garcia-Calderon and J. G. Muga, Phys. Rep. {\bf 476}, 1 (2009).
\bibitem{Caldeira1983} A. O. Caldeira and A. J. Leggett, Annals of Physics {\bf 149}, 374 (1983).
\bibitem{Bruinsma1986} R. Bruinsma and P. Bak, Phys. Rev. Lett. {\bf 56}, 420 (1986).
\bibitem{Buttiker1982} M. B\"{u}ttiker and R. Landauer, Phys. Rev. Lett. {\bf 49}, 1739 (1982).
\bibitem{Hauge1989} E. H. Hauge and J. A. Stovneng, Rev. Mod. Phys. {\bf 61}, 917 (1989); E. Eisenberg and Y. Ashkenazy, Found. Phys.
 {\bf 27}, 191 (1997); A. V. Pimpale, Prog. Quant. Elect. {\bf 28}, 345 (2004).
\bibitem{Switkes1999} M. Switkes, C. M. Marcus, K. Campman and A. C. Gossard, Science {\bf 289}, 1905 (1999); S. Kohler, J. Lehmann and P. Haenggi, Phys. Rep. {\bf 406}, 379 (2005).
\bibitem{Miller1987} D. A. B. Miller, Optical Engineering {\bf 26}, 368 (1987); N. H. Bonadeo {\it et al}, Science {\bf 282}, 1473 (1998); F. H. L. Koppens {\it et al}, Nature {\bf 442}, 766 (2006).
\bibitem{Pimpale1991} A. Pimpale and M. Razavy, Fortschritte der Physik/Progress of Physics {\bf 39}, 85 (1991); E. Cota, J. V. Jose and F. Rojas, Nanostructured Materials {\bf 3}, 349 (1993); A. V. Pimpale, Progress in Quantum Electronics {\bf 28}, 345 (2004).
\bibitem{Bagwell1992} P. F. Bagwell and R. K. Lake, Phys. Rev. {\bf B 46}, 15329 (1992).
\bibitem{Martinez2001} D. F. Martinez and L. E. Reichl, Phys. Rev. {\bf B 64}, 245315 (2001).
\bibitem{Moskalets2003} M. Moskalets and M. B\"{u}ttiker, Phys. Rev. {\bf B 68}, 075303 (2003); S. W. Kim, Int. J. Mod. Phys. {\bf B 18}, 3071 (2004); C. Benjamin, Eur. Phys. J. {\bf B 52}, 403 (2006); M. M. Mahmoodian, L. S. Braginskii and M. V. Entin, Phys. Rev. {\bf B 74}, 125317 (2006); M. M. Mahmoodian and M. V. Entin, EPL {\bf 77}, 67002 (2007);
M. M. Makhmudian, M. V. Entin and L. S. Braginskii, Journal of Experimental and Theoretical Physics {\bf 105}, 495 (2007);
M. Moskalets and M. B\"{u}ttiker, Phys. Rev. {\bf B 75}, 035315 (2007).
\bibitem{Diakonos2011} F. K. Diakonos, P. Kalozoumis, A. I. Karanikas and P. Schmelcher, in preparation.
\bibitem{Martinez2002} D. F. Martinez, L. E. Reichl and G. A. Luna-Acosta, Phys. Rev. {\bf B 66}, 174306 (2002).
\bibitem{Tannor2007} D. J. Tannor, {\it{``Introduction to quantum mechanics, a time-dependent perspective"}}, University Science Books, 2007.
\bibitem{Kleinert2009} H. Kleinert, {\it{``Path Integrals in Quantum Mechanics, Statistics, Polymer Physics, and Financial Markets"}},
World Scientific, Singapore (2009).
\bibitem{Storchak1992} S. N. Storchak, Phys. Lett. {\bf A 161}, 397 (1992); V. V. Dodonov, V. I. Manko and D. E. Nikonov, Phys. Lett. {\bf A 162}, 359 (1992); C. Grosche, Phys. Lett. {\bf A 182}, 28 (1993); S. Albeverio and S. Mazzuchi, Journal of Functional Analysis {\bf 238}, 471 (2006); D. Laroze, G. Gutierez, R. Rivera and J. M. Yanez,
Phys. Scr. {\bf 78}, 015009 (2008); S. Pepore and B. Sukbot, Indonesian Journal of Physics {\bf 21}, 47 (2010).
\bibitem{Balaz2011} A. Balaz {\it{et al}}, J. Stat. Mech. (2011) P03004; ibid P03005.
\bibitem{Campbell2009} J. Campbell, J. Phys. {\bf A 42}, 365212 (2009).
\bibitem{Fujikawa2008} K. Fujikawa, Phys. Rev. {\bf D 77}, 045006 (2008).
\end{thebibliography}
\end{document}